\documentstyle{article}

\newskip\humongous \humongous=0pt plus 1000pt minus 1000pt

\newif\ifdtup

\def\oldreffmt#1{\rlap{[#1]} \hbox to 2\parindent{}}

\def\figfmt#1{\rlap{Figure {#1}} \hbox to 1in{}}

\def\beq{\begin{equation}}
\def\eeq{\end{equation}}

\relax

\font\tenbf=cmbx10
\font\tenrm=cmr10
\font\tenit=cmti10
\font\elevenbf=cmbx10 scaled\magstep 1
\font\elevenrm=cmr10 scaled\magstep 1
\font\elevenit=cmti10 scaled\magstep 1

\font\ninerm=cmr9

\renewenvironment{thebibliography}[1]
 { \elevenrm
   \begin{list}{\arabic{enumi}.}
    {\usecounter{enumi} \setlength{\parsep}{0pt}
     \setlength{\itemsep}{3pt} \settowidth{\labelwidth}{#1.}
     \sloppy
    }}{\end{list}}

\begin{document}
\begin{center}{{\tenbf FERMION MASSES IN SO(10)\footnote{\ninerm Presented
at DPF 92 Meeting, Fermilab, November, 1992.}\\}
\vspace{-1in}
\rightline{EFI 92-64}
\rightline{November 1992}
\rightline{hep-ph/9301248}
\bigskip
\vglue 1.8cm
{\tenrm GERARD JUNGMAN \\}
\baselineskip=12pt
{\tenit Enrico Fermi Institute, University of Chicago, 5640 S. Ellis Ave.\\}
\baselineskip=12pt
{\tenit Chicago,IL 60637, USA\\}
\vglue 0.3cm
\vglue 0.8cm
{\tenrm ABSTRACT}}
\end{center}
\vglue 0.2cm
{\rightskip=3pc
 \leftskip=3pc
 \tenrm\baselineskip=11pt
 \noindent
%
%
Yukawa coupling constant unification together with the known fermion
masses is used to constrain SO(10) models. We consider the case of
one (heavy) generation, with the tree-level relation $m_b=m_\tau$,
calculating the limits on the
intermediate symmetry breaking scales.
This analysis extends previous analyses which addressed only
the simplest symmetry breaking schemes.
In the case where the low energy model is the standard model
with one Higgs doublet, there are very strong constraints
due to the known limits on the top mass and the tau neutrino
mass. The two Higgs doublet case is less constrained.
Finally we address the role of
a speculative constraint on the tau neutrino mass, arising
from the cosmological implications of anomalous $B+L$ violation
in the early universe.

\vglue 0.6cm}
{\elevenbf\noindent 1. Motivation}
\elevenrm
\baselineskip=13.8pt

Non-zero neutrino masses arguably provide one of the more well motivated
extensions of the standard model. Theoretically, there is a
prejudice for a see-saw \cite{seesaw} neutrino mass mechanism since the
relative smallness of the neutrino masses is thereby naturally explained.
Such a mechanism is readily available in grand-unified theories which
display spontaneous violation of lepton number at a high scale.
The simplest such model is the $SO(10)$ GUT \cite{soten}.
Of the two possible maximal subgroups which can appear in the symmetry
breaking chain, only $SU(2)_L\times SU(2)_R \times SU(4)$ is viable
phenomenologically. This is the Pati-Salam \cite{PatSal}
intermediate unification,
and displays the left-right symmetry directly. In many ways, $SO(10)$ is
the canonical implementation of grand unification with spontaneous
violation of lepton number. Thus it seems worthwhile to study neutrino
masses in this model, in particular their relation to the other fermion
masses so that we can understand the constraints on the model from measured
(or constrained) masses.

Two major features of the current experimental situation have
led to a revival in GUT calculations.
First, there is the increasing precision
of coupling constant measurements at the $Z$ resonance.
Second, there is the increasing lower bound on the top quark mass.
Precision measurements of gauge couplings provide constraints
on intermediate symmetry breaking scales, and a large $m_t$-$m_b$
splitting may be difficult to reconcile with a unified fermion multiplet
without some gymnastics. Furthermore, such a large splitting has important
consequences for neutrino masses, since the lepton sector must
mirror the quark sector.

The first step in gaining a quantitative understanding of fermion masses
in $SO(10)$ is to calculate radiative corrections to the tree-level
mass relations within the fermion multiplet. The results of
these calculations and their implications are discussed in Ref. \cite{me}.

Considering the nature of this note, references are sparse. For a more
complete list of references, we refer to Ref. \cite{me}.

\vglue 0.4cm
{\elevenbf\noindent 2. The Model }

$SO(10)$ has nothing to say about the repetition of generations exhibited
by nature,
and as a first approximation
we deal only with the heavy generation. One generation of fermions
can Yukawa couple only to the scalar representations $\phi({\bf 10})$ or
$\phi({\bf 126})$,
and it must have Dirac Yukawa couplings predominantly \cite{metwo}
to the low-energy Higgs doublets which derive from $\phi({\bf 10})$. This
gives the tree-level relations $m_b=m_\tau$ and $m_t=m_{\nu_\tau}^{\rm Dirac}$.
Note that a small admixture of $\phi({\bf 126})$ mass relations
is required in order to produce the correct pattern
of mixing in the full multi-generation model, but this can arise
in ways which do not require a finely tuned upset to these tree-level
relations \cite{Babu}.

Of course, we must realize that there is a great deal of freedom for
engineering the fermion spectrum of the model by introducing
arbitrary complications in the Higgs sector.
Therefore, before making predictions, we must
choose a philosophy for a minimal model. This is less necessary when
dealing with only one generation than it is when dealing with a full
multi-generation model, considering the above remarks about couplings.
We choose
the minimal model to have only the Yukawa coupling of $\phi({\bf 10})$
necessary to generate the tree-level Dirac masses and the Yukawa
coupling of $\phi({\bf 126})$ necessary to generate the tree-level
Majorana mass for the right-handed neutrino.
Our philosophy is to constrain the minimal model as much as possible.

The last piece of information necessary to specify the model is the number
of Higgs doublets that we wish to have in the low-energy theory.
Derived from a single $\phi({\bf 10})$, there can be one or two
Higgs doublets. This choice is important for the radiative corrections,
embodied by the Yukawa coupling renormalization group equations. In the
one doublet model, the large $m_t$-$m_b$ splitting requires a large
splitting in the top and bottom Yukawa couplings. In the two doublet
model, the mass splitting is accounted for by the large ratio of the
two Higgs doublet vevs, $\tan\beta = v_u/v_d \simeq 30$.

The Yukawa coupling beta functions are detailed  in Ref. \cite{me}.

\vglue 0.4cm
{\elevenbf\noindent 3. Results }

The constraints we place on the model are as follows. First, the
gauge coupling unification must be consistent with the measured
values of gauge couplings; $\alpha_S(M_Z) = 0.11\pm 0.01$,
$\sin^2\theta_W = 0.233\pm 0.003$, and $\alpha(M_Z) = 0.00781$ with
negligible error for our analysis. Second, the fermion masses must
be consistent with $m_\tau=1.78\;{\rm GeV}$, $m_t > 91 \;{\rm GeV}$,
and $m_b = 4.3\pm 0.3 \;{\rm GeV}$. The latter of these is clearly an important
parameter, and an  attempt is made in Ref. \cite{me} not to hide the
dependence on $m_b$. Remember this value of $m_b$ is the current
mass and not the constituent mass which is fit in potential models
to be $m_b^{Cons}\simeq 4.8 \;{\rm GeV}$. Also note that we use a value of
$\alpha_S(M_Z)$ which is somewhat smaller than the direct determinations.
Such a value can arise from certain analyses \cite{Ellis}.
This is a conservative assumption for our purposes, as will be explained
below. Finally, we apply the weakest of the cosmological constraints
on the $\nu_\tau$ mass \cite{cowsik} together with the direct limit,
$m_{\nu_\tau} < 70 {\rm eV} \;or\; 1 {\rm MeV} < m_{\nu_\tau} < 35 {\rm MeV}$.

Consider first the case of one low-energy Higgs doublet. Previously,
it had been shown that it was difficult to reconcile the large
$m_t$-$m_b$ splitting with unification and the measured $m_b/m_\tau$ ratio,
in a restricted symmetry
breaking scheme \cite{bartolguys}. By relaxing the constraint
on the symmetry breaking scheme, we allow the top quark a bit more range,
but the basic conclusion remains the same. For $\alpha_S(M_Z) > 0.105$
the one Higgs doublet case is ruled out at the one sigma level, unless
the tau neutrino mass lies in its upper window. Larger values of
$\alpha_S(M_Z)$ produce an unacceptably large $m_b/m_\tau$, and thus
we see why our choice of ``smaller'' values for $\alpha_S(M_Z)$ is
a conservative one.

The case of two low-energy Higgs doublets is almost unconstrained
by the $m_b/m_\tau$ ratio. This is because the beta function for the
bottom quark Yukawa coupling is slightly more positive in this case,
and $m_b$ evolves into the middle of its allowed range.

Finally, we come to what may be the strictest of the constraints on
these models. This is the constraint due to the observed baryon asymmetry
of the universe. If lepton-number violating processes are in equilibrium
with anomalous $B+L$ violating processes in the early universe,  the
baryon asymmetry will be washed away \cite{hartur}\cite{fukyan}.
At tree-level this constrains $m_{\nu_\tau}$ sufficiently that the model
{\elevenit cannot support} the known lower bound on the top mass, with no
regard to
details such as the number of low-energy Higgs doublets. However, we find
that the radiative corrections are large enough that this conclusion
is tempered to a bound (roughly) $m_t < 120 \;{\rm GeV}$. The real test
of this sort of argument remains to be completed. The question
must be addressed in
the light of a full leptogenesis calculation, incorporating CP violation
in the lepton sector and realistic multi-generation mass matrices \cite{menow}.

\vglue 0.4cm
{\elevenbf\noindent 4. References \hfil}
\vglue 0.01cm


\begin{thebibliography}{9}
\bibitem{seesaw} M.~Gell-Mann, P.~Ramond, and R.~Slansky, in {\elevenit
Supergravity}, North Holland (1979)
\bibitem{soten} H.~Georgi and D.~Nanopoulos, {\elevenit Nucl.~Phys.} {\elevenbf
B159} (1979) 16.
\bibitem{PatSal} J.~C.~Pati and A.~Salam, {\elevenit Phys.~Rev.} {\elevenbf
D10} (1974) 275.
\bibitem{me} G.~Jungman, {\elevenit Phys.~Rev.} {\elevenbf D46} (1992) 4004.
\bibitem{metwo} G.~Jungman, {\elevenit Phys.~Rev.} {\elevenbf D} Brief Report,
to appear.
\bibitem{Babu} K.S.~Babu and R.N.~Mohapatra, Bartol preprint BA-92-054 (1992).
\bibitem{Ellis} J.~Ellis, D.~V.~Nanopoulos, and D.~A.~Ross, preprint
CERN-TH.6130/91 (1991).
\bibitem{cowsik} H.~Harari and Y.~Nir, {\elevenit Nucl.~Phys.} {\elevenbf B292}
 (1987) 251.
\bibitem{bartolguys} G.~Lazarides and Q.~Shafi, {\elevenit Nucl.~Phys.}
{\elevenbf B350} (1991) 179; E.~M.~Freire, {\elevenit Phys.~Rev.} {\elevenbf
D43} (1991) 209.
\bibitem{hartur} J.~Harvey and M.~Turner, {\elevenit Phys.~Rev.} {\elevenbf
D42} (1990) 3344.
\bibitem{fukyan} M.~Fukugita and T.~Yanagida, {\elevenit Phys.~Rev.} {\elevenbf
D42} (1990) 1285.
\bibitem{menow} G.~Jungman, work in progress.
\end{thebibliography}
\end{document}